\documentstyle[11pt,fullpage]{article}

\def\half{\textstyle{1 \over 2}}
\def\01{\{0,1\}}
\def\x{\times}
\def\ox{\otimes}
\def\xor{\oplus}
\def\Q{\mbox{\sf Q}}
\def\ket#1{\mbox{$| #1 \rangle$}}

\def\ni{\noindent}

\def\ee{\vspace*{2mm}}

\def\loud#1{\noindent{\bf #1 }}

\begin{document}

\title{Quantum Entanglement and Communication Complexity}

\author{Harry Buhrman\,%
\thanks{Partially supported by: 
NWO by  SION Project 612-34-002, 
EU through NeuroCOLT ESPRIT Working Group Nr.\ 8556, 
and HC\&M grant CCR 92-09833.} \\
{\protect\small\sl CWI, Amsterdam\/}\,%
\thanks{CWI, Kruislaan 413, 1098 SJ Amsterdam, The Netherlands.
E-mail: {\tt buhrman@cwi.nl}, {\tt wimvdam@cwi.nl}.}
\and
Richard Cleve\,%
\thanks{Partially supported by Canada's NSERC.} \\
{\protect\small\sl University of Calgary\/}\,%
\thanks{Department of Computer Science, University of Calgary, Calgary, 
Alberta, Canada T2N 1N4.
E-mail: {\tt cleve@cpsc.ucalgary.ca}.}
\and
Wim van Dam \\
{\protect\small\sl CWI, Amsterdam\/}\,\footnotemark[2]
}

\date{}

\maketitle

\begin{abstract}
We consider a variation of the multi-party communication complexity 
scenario where the parties are supplied with an extra resource: 
particles in an entangled quantum state.
We show that, although a prior quantum entanglement cannot be used 
to simulate a communication channel, it can reduce the communication 
complexity of functions in some cases.
Specifically, we show that, for a particular function among three parties 
(each of which possesses part of the function's input), a prior quantum 
entanglement enables them to learn the value of the function with 
only three bits of communication occurring among the parties, whereas, 
without quantum entanglement, four bits of communication are necessary.
We also show that, for a particular two-party probabilistic communication 
complexity problem, quantum entanglement results in less communication 
than is required with only classical random correlations (instead of quantum 
entanglement).
These results are a noteworthy contrast to the well-known fact that 
quantum entanglement cannot be used to actually simulate communication 
among remote parties.
\end{abstract}


\section{Introduction and summary of results}

One of the most remarkable aspects of quantum physics is the notion 
of {\em quantum entanglement}.
If two particles are in an entangled state then, even if the particles 
are physically separated by a great distance, they behave in some 
respects as a single entity rather than as two separate entities.
The entangled particles exhibit what physicists call {\em nonlocal\/} 
effects.
Informally, these are effects that cannot occur in a world governed by 
the laws of ``classical'' physics unless communication occurs between 
the particles.
Moreover, if the physical separation between the particles is large and 
the time between the observations is small then this entailed 
communication may exceed the speed of light!
Nonlocal effects were alluded to in a famous 1935 paper by Einstein, 
Podolsky, and Rosen \cite{EPR}.
Einstein later referred to this as {\em spukhafte Fernwirkungen} 
[spooky actions at a distance] (see \cite{BE71,Mer85} for more 
historical background).
In 1964, Bell \cite{Bell} formalized the notion of two-particle 
nonlocality in terms of correlations among probabilities in a scenario 
where one of a number of a measurements are performed on each particle.
He showed that the results of the measurements that occur quantum 
physically can be correlated in a way that cannot occur classically 
unless the type of measurement selected to be performed on one particle 
affects the result of the measurement performed on the other particle.

In reality---which is quantum physical---the nonlocal effects exhibited 
by entangled particles do not involve any communication (consequently, 
nonlocality does not entail communication faster than the speed of light).
In operational terms, the ``spooky actions at a distance'' that Einstein 
referred to cannot be used to simulate a communication channel.
More precisely, if two physically separated parties, Alice and Bob, 
initially possess entangled particles and then Alice is given an arbitrary 
bit $x$, there is no way for Alice to manipulate her particles in order 
to convey $x$ to Bob when he performs measurements on his particles.
The probabilities pertaining to any conceivable measurement that 
Bob can perform on his particles are all determined by the {\em (reduced) 
density matrix} of Bob's particles (see \cite{Peres} for definitions 
and discussion), and this density matrix does not change when Alice 
manipulates her particles by unitary transformations and measurements.
Moreover, entanglement cannot even be used to {\em compress} information 
in the following sense: for Alice to convey $n$ arbitrary bits to Bob, 
she must in general send $n$ bits---sending $n-1$ bits will not suffice.

Similar results apply to communications involving more than two 
parties.
For example, suppose that Alice, Bob, and Carol share entangled particles, 
and then each is given an arbitrary $n$-bit string.
If each party wants to convey his $n$ bits to the other parties 
using (say) a global broadcast channel then, in spite of the quantum 
entanglement, Alice must send $n$ bits to the channel, and so must 
Bob and Carol.
The argument is again in terms of the fact that the reduced density 
matrix of each party's particles cannot be changed by the other parties.

Now, consider the related but different scenario of 
{\em communication complexity}.
Yao \cite{Yao} introduced and investigated the following problem.
Alice obtains an $n$-bit string $x$, and Bob obtains an $n$-bit string $y$ 
and the goal is for both of them to determine $f(x,y)$, for some function 
$f : \01^n \x \01^n \rightarrow \01$, 
with as little communication between them as possible.
Clearly, $n+1$ bits of communication always suffice (Alice sends all 
her $n$ bits to Bob, Bob computes $f(x,y)$, and sends the one-bit answer 
to Alice), but for some functions fewer bits suffice.
This scenario and variations of it have been widely studied (see 
\cite{KN} for an extensive survey).

In one variation of the above communication complexity scenario, 
there are more than two parties, each of which is given a subset of 
the input data.
In another variation, all parties have access to a common string of 
random bits.
This string can be assumed to have been communicated during a ``set up'' 
stage, prior to the parties' being given their input data.
For some functions, this prior random string reduces the communication 
complexity for a worst-case input if a small error probability is 
permitted (here, a ``worst-case input'' is understood to be chosen 
independently of the random string).
If no error probability is allowed then a prior shared random string 
does not reduce the communication complexity (in a worst-case execution).
Also, if the input is selected randomly with respect to some arbitrary 
but fixed probability distribution then, even if a particular error 
probability is permitted, the communication complexity does not decrease 
by having a prior shared random string (\cite{Yao,KN}).

In the present paper, we consider a variation of the above ``classical'' 
communication complexity scenarios where a prior quantum entanglement 
is available to the parties.
We first consider the case where no error probability is permitted.
On the face of it, it may appear that a prior quantum entanglement 
does not reduce the communication complexity of functions.
Consider the following informal argument, which we phrase in a 
three-party setting, where Alice, Bob, and Carol are given input 
strings $x$, $y$, and $z$ respectively, and the goal is to collectively 
determine $f(x,y,z)$:
\begin{enumerate}
\item 
Assume that the classical communication complexity of function $f(x,y,z)$ 
is $k$.
That is, $k$ bits of communication are necessary for Alice, Bob, and Carol 
to acquire the answer.
\item
A prior entanglement cannot simulate or even compress any particular 
act of communication.
\item
{\em Ergo}, even with a prior quantum entanglement, the communication 
complexity of function $f(x,y,z)$ is $k$.
\end{enumerate}
We shall demonstrate that this conclusion is incorrect by a 
counterexample $f(x,y,z)$ where: 
without a prior quantum entanglement, four bits of 
communication are {\em necessary} to compute $f(x,y,z)$; whereas, 
with a prior quantum entanglement, three bits of 
communication are {\em sufficient} to compute $f(x,y,z)$.

Our protocol employing quantum entanglement uses less communication than 
necessary by any classical protocol by manipulating entangled states to 
{\em circumvent\/} (rather than simulate) communication.
Our technique is based on an interesting variation of Bell's Theorem, due 
to Mermin \cite{Mermin}, for three particles which is ``deterministic'' 
in the sense that all the associated probabilities are either zero or one.
Mermin's result is a refinement of a previous four-particle result,  
due to Green\-berger, Horne, and Zeilinger \cite{GHZ}.

We also give an example of a two-party probabilistic communication 
complexity scenario with a function $g(x,y)$ where: 
with a classical shared random string but no prior quantum entanglement, 
three bits of communication are {\em necessary} to compute $g(x,y)$ with 
probability at least $\cos^2({\pi \over 8}) = 0.853...$; whereas, 
with a prior quantum entanglement, two bits of communication are 
{\em sufficient} to compute $g(x,y)$ with the same 
probability.
This is based on a variation of Bell's Theorem, due to Clauser, Horne, 
Shimony, and Holt \cite{CHSH}.

Although, in both of the above cases, the savings in communication are 
not in an asymptotic setting, we consider these results as evidence that 
quantum entanglement can potentially change the nature of communication 
complexity.

This paper is an extension of our previous version of these results 
\cite{CB}.
Also, Grover \cite{Grover} recently independently obtained results 
related to communication complexity in a probabilistic setting.

\section{A three-party deterministic scenario}

Consider the following three-party scenario.
Alice, Bob, and Carol receive $x$, $y$, and $z$ respectively, 
where $x,y,z \in \{0,1,2,3\}$, and the condition 
\begin{equation}
x+y+z \equiv 0 \pmod{2}
\label{promise}
\end{equation}
is promised.
The common goal is to compute the value of the function 
\begin{equation}
f(x,y,z) = \frac{(x+y+z) \bmod{4}}{2},
\label{modfoursum}
\end{equation}
(which has value 0 or 1 by Eq.~(\ref{promise})).
We represent the numbers $x$, $y$, and $z$ in binary notation 
as $x_1 x_0$, $y_1 y_0$, and $z_1 z_0$.
In terms of these bits, the promise of Eq.~(\ref{promise}) is 
\begin{equation}
x_0 \xor y_0 \xor z_0 = 0, 
\label{promise2}
\end{equation}
and the function of Eq.~(\ref{modfoursum}) for inputs satisfying 
Eq.~(\ref{promise2}) is 
\begin{equation}
f(x,y,z) = x_1 \xor y_1 \xor z_1 \xor (x_0 \vee y_0 \vee z_0).
\label{modfoursum2}
\end{equation}
We assume the standard multi-party communication channel where 
each bit that a party sends is broadcast to all other parties.
Also, at the conclusion of the protocol, {\em all\/} parties must 
be able to determine the value of the function.

In the following two subsections, we show that, with a prior quantum 
entanglement, three bits of communication are sufficient to compute 
$f(x,y,z)$, whereas, without a prior quantum entanglement, four bits of 
communication are necessary to compute $f(x,y,z)$.

\subsection{The communication complexity with quantum entanglement is 
three bits}

We now show that if Alice, Bob, and Carol initially share a
certain entanglement of three qubits then there is a protocol in
which each party broadcasts one classical bit such that 
the value $f(x,y,z)$ is known to all parties afterwards. 
The entanglement is 
\begin{equation}
\ket{\Q_A \Q_B \Q_C} = 
\half({|000\rangle - |011\rangle - |101\rangle - |110\rangle}),
\label{state}
\end{equation} 
where Alice, Bob, and Carol have qubits $\Q_A$, $\Q_B$, and $\Q_C$, 
respectively.
(This is equivalent to the state examined in \cite{Mermin} in an 
alternate basis.)

The idea is based on applying Mermin's result \cite{Mermin} to enable 
Alice, Bob, and Carol to obtain bits $a$, $b$, and $c$ respectively, 
such that $a \xor b \xor c = x_0 \vee y_0 \vee z_0$.
This is achieved by the following procedures:\ee
\begin{tabbing}
\hspace*{8mm} Pr\=ocedure for Alice: \hspace*{5cm} Pr\=ocedure for Bob: \\
\> {\bf if} $x_0 = 1$ {\bf then} {\bf apply $H$ to} $\Q_A$ 
\> {\bf if} $y_0 = 1$ {\bf then} {\bf apply $H$ to} $\Q_B$ \\
\> {\bf measure} $\Q_A$ {\bf yielding bit $a$} 
\> {\bf measure} $\Q_B$ {\bf yielding bit $b$} \\
\end{tabbing}
\begin{tabbing}
\hspace*{8mm} Pr\=ocedure for Alice: \hspace*{5cm} Pr\=ocedure for Bob: \kill
\hspace*{8mm} Pr\=ocedure for Carol: \\
\> {\bf if} $z_0 = 1$ {\bf then} {\bf apply $H$ to} $\Q_C$ \\
\> {\bf measure} $\Q_C$ {\bf yielding bit $c$} \\
\end{tabbing}
In the above, $H$ is the Hadamard transform, which is represented 
in the standard basis ($\ket{0}$ and $\ket{1}$) as 
\begin{equation}
H = \textstyle{\frac{1}{\sqrt{2}}}
\pmatrix{1 & \ \ 1 \cr 1 & -1},
\end{equation}
and the measurements are performed in the standard basis.
In the language of operators, if a party's low-order bit 
is 0 then the measurement is with respect to the Pauli 
matrix $\sigma_z$; otherwise (if the low-order bit is 1), 
the measurement is with respect to the Pauli matrix $\sigma_x$.\ee

\loud{Lemma 1:}{\sl 
In the described procedure, $a \xor b \xor c = x_0 \vee y_0 \vee z_0$.}\ee

\loud{Proof:} 
Recall that, by the promise of Eq.~(\ref{promise2}), 
$x_0 y_0 z_0 \in \{000,011,101,110\}$.

First, consider the case where $x_0 y_0 z_0 = 000$.
In this case, no $H$ transform is applied to any of the 
qubits $\Q_A$, $\Q_B$, or $\Q_C$.
Therefore, $\Q_A \Q_B \Q_C$ is measured in state (\ref{state}), which 
implies that $a \xor b \xor c = 0 = x_0 \vee y_0 \vee z_0$.

Next, in the case where $x_0 y_0 z_0 = 011$, 
a Hadamard transform is applied to $\Q_B$ and to $\Q_C$, but not to $\Q_A$.
Therefore, $\Q_A \Q_B \Q_C$ is measured in state 
\begin{eqnarray}
I \ox H \ox H 
\left({\half ({|000\rangle -|011\rangle -|101\rangle -|110\rangle})}\right) 
& = & \half ({|001\rangle+|010\rangle-|100\rangle+|111\rangle}), 
\end{eqnarray}
so 
$a \xor b \xor c = 1 = x_0 \vee y_0 \vee z_0$.
The remaining cases where $x_0 y_0 z_0 = 101$ and $110$ are similar 
by the the symmetry of state (\ref{state}).
$\Box$\ee

After the above procedure, Alice broadcasts the bit $(x_1 \xor a)$, 
Bob broadcasts $(y_1 \xor b)$, and Carol broadcasts $(z_1 \xor c)$.
At this point, each party knows $(x_1 \xor a)$, $(y_1 \xor b)$, 
and $(z_1 \xor c)$, from which they can each determine the bit 
\begin{eqnarray}
(x_1 \xor a) \xor (y_1 \xor b) \xor (z_1 \xor c) 
& = & x_1 \xor y_1 \xor z_1 \xor (a \xor b \xor c) \nonumber \\
& = & x_1 \xor y_1 \xor z_1 \xor (x_0 \vee y_0 \vee z_0), 
\end{eqnarray}
as required.

\subsection{The communication complexity without quantum entanglement is 
four bits}
\label{section:lowerbound} 

In this section, we show that, in the classical setting, four bits of
communication are necessary to compute $f(x,y,z)$.

One can view any $k$-bit protocol as a binary tree of depth $k$, where 
each node that is not a leaf is labelled A(lice), B(ob), or C(arol).
This labelling indicates which party will broadcast the next bit.
An execution of the protocol corresponds to a path from the root of 
the tree to a leaf.
Each leaf node is labelled 0 or 1, indicating the common output that 
results from the execution leading to that leaf.
To establish our lower bound, it suffices to show that no protocol-tree 
of depth three correctly computes $f(x,y,z)$.

We use the following lemma, which implies that, in any correct protocol, 
all three parties must broadcast at least one bit.\ee

\loud{Lemma 2:}{\sl For any correct protocol-tree, on every path from its 
root to a leaf, each of\/} A, B, {\sl and\/} C {\sl must occur as a label 
at least once.}\ee

\loud{Proof:}
Suppose that there exists a path along which one party, say A, does not 
occur as a label.
Let the leaf of that path be labelled $l \in \01$.
Since this path does not include any reference to Alice's data, the same 
path is taken if $x_1$ is negated and all other input bits are held constant.
But, by Eq.~(\ref{modfoursum2}), negating $x_1$ also negates the value 
of $f(x,y,z)$, so the protocol cannot be correct for both possible 
values of $x_1$.
$\Box$\ee

Next, suppose we have a protocol-tree of depth three for $f(x,y,z)$.
Assume, without loss of generality, that the root of the tree is 
labelled A.
The bit that Alice broadcasts is some function 
$\phi : \01^2 \rightarrow \01$ of her input data $x$ alone.
The function $\phi$ partitions $\01^2$ into two classes $\phi^{-1}(0)$ 
and $\phi^{-1}(1)$.
Call these two classes $S_0$ and $S_1$, and assume (without loss of 
generality) that $00 \in S_0$.

Next, assume for the moment that the two children of the root of 
the protocol-tree are both labelled B (we shall see later that the other 
cases can be handled similarly).
Then, by Lemma~2, the four children of B are all labelled C.
Therefore, after Alice and Bob each send a bit, Carol must have enough 
information to determine the value of $f(x,y,z)$, since Carol broadcasts 
the third bit and does not gain any information from doing this.
We shall show that this is impossible whatever $S_0$ and $S_1$ are.

There are two cases (the second of which has three subcases):\ee

\loud{Case 1 {\boldmath $|S_0| = \mbox{\bf 1}$}:}
Recall that $00 \in S_0$, so $01, 10, 11 \in S_1$.
Now, should the bit that Alice broadcasts specify that $x \in S_1$, 
Bob must follow this by broadcasting one bit from which Carol can 
completely determine the value of $f(x,y,z)$.
Suppose that Bob sends the bit consistent with $y = 01$.
If $z = 00$ then, from Carol's perspective, the possible values of 
$(x,y,z)$ include $(01,01,00)$ and $(11,01,00)$, for which the respective 
values of $f(x,y,z)$ are $1$ and $0$.
Therefore, Carol cannot determine the value of $f(x,y,z)$ in this case.\ee

\loud{Case 2 {\boldmath $|S_0| \ge \mbox{\bf 2}$}:}
There are three subcases where $S_0$ contains $01$, $10$ or $11$, 
in addition to $00$.\ee

\loud{Case 2.1 {\boldmath $S_0$ contains 00 and 01}:}
Here, we consider the case where Alice broadcasts the bit specifying that 
$x \in S_0$.
Bob must follow this by broadcasting one bit from which Carol can 
completely determine the value of $f(x,y,z)$.
The bit that Bob broadcasts induces a  partition of the possible values for 
$y$ into two classes.
If $z = 00$ then, from Carol's perspective, after receiving Alice's bit 
but before receiving Bob's bit, the possible values of $(x,y,z)$ 
include $(00,00,00)$, $(00,10,00)$, $(01,01,00)$, and $(01,11,00)$, 
and the respective values of $f(x,y,z)$ on these points are $0$, $1$, $1$, 
and $0$.
Therefore, for the protocol to be successful in this case, 
the partition that Bob's bit induces on $y$ must place $00$ and $11$
in one class and $01$ and $10$ in the other class (otherwise Carol would not 
be able to determine $f(x,y,z)$ when $z = 00$).
On the other hand, if $z = 01$ then, from Carol's perspective, the possible 
values of $(x,y,z)$ include $(00,01,01)$, $(00,11,01)$, $(01,00,01)$, and
$(01,10,01)$ and the respective values of $f(x,y,z)$ on these points are 
$1$, $0$, $1$, and $0$.
Since we have established that Bob's bit does not distinguish between 
$y = 00$ and $y = 11$, Bob's bit is not sufficient information for Carol 
to determine $f(x,y,z)$ in this case.\ee

\loud{Case 2.2 {\boldmath $S_0$ contains 00 and 10}:}
The argument is similar to that in Case 1.
Assume that Alice sends the bit specifying that $x \in S_0$.
If Bob follows this by sending the bit consistent with $y = 00$ and 
$z = 00$ then, from Carol's perspective, the possible values of 
$(x,y,z)$ include $(00,00,00)$ and $(10,00,00)$ and the respective 
values of $f(x,y,z)$ on these points are $0$ and $1$.
Thus, Carol cannot determine the value of $f(x,y,z)$ in this case.\ee

\loud{Case 2.3 {\boldmath $S_0$ contains 00 and 11}:}
The argument is similar to Case 2.1.
Suppose that Alice broadcasts the bit specifying that $x \in S_0$.
Consider Carol's perspective.
If $z = 00$ then the possible values of $(x,y,z)$ include 
$(00,00,00)$, $(00,10,00)$, $(11,01,00)$, and $(11,11,00)$
and the respective values of $f(x,y,z)$ on these points are 
$0$, $1$, $0$, and $1$; whereas, if $z = 01$ then the possible values
of $(x,y,z)$ include $(00,01,10)$, $(00,11,01)$, $(11,00,01)$, and
$(11,10,01)$ and the respective values of $f(x,y,z)$ on these points are 
$1$, $0$, $0$, $1$.
No binary partitioning of $y$ will work for both possibilities.\ee

The cases were the two children of the root of the protocol-tree are CC, 
CB and BC have an analogous proof as above with the roles of B and C 
possibly reversed.

This completes the proof of the lower bound of four bits.
There is a straightforward four-bit protocol, demonstrating that 
this bound is tight.

\subsection{Application to a three-party variation of the inner product 
function}

In this section, we show that $f(x,y,z)$ is a generalization of a 
restricted version the three-party inner product.
The following function is considered in \cite{CB}.
Alice, Bob, and Carol are given $n$-bit strings $x$, $y$, and $z$ 
respectively, which are subject to the condition that 
\begin{equation}
x \xor y \xor z = \overbrace{11 \ldots 1}^n,
\label{conds}
\end{equation}
(where $\xor$ is applied bitwise) and the goal is to determine the function 
\begin{equation}
\mbox{\it GIP\/}(x,y,z) = 
(x_1 \wedge y_1 \wedge z_1) \xor \cdots \xor (x_n  \wedge y_n  \wedge z_n).
\label{fcn}
\end{equation}
An alternative way of expressing this problem is to impose no restriction 
on the inputs, $x$, $y$, $z$, and to extend $GIP$ to a {\em relation} such 
that on the points where Eq.~(\ref{conds}) is violated, both 0 and 1 are 
acceptable outputs.
Clearly, this problem has the same communication complexity as the 
original one.

Note that, from the perspective of any two of the three parties, 
this problem is exactly equivalent to the two-party inner product.
Thus, if only two parties participate in the communication, 
the classical communication complexity is the same as that of the 
two-party inner product function, which is $n+1$ (\cite{KN}).
{}From this, one might suspect that, even if all three parties participate 
in the communication, the classical communication complexity remains 
close to $n$.
In fact, in \cite{CB}, it is shown that, to solve this problem, it 
suffices for Alice, Bob, and Carol to: 
(a) count the number of 0s in their respective input strings; 
(b) determine the sum of these three quantities modulo four.
Also, this sum must be even.
This is equivalent to the problem defined by Eqs.~(\ref{promise}) 
and (\ref{modfoursum}), which has classical communication complexity 
four, and quantum communication complexity three.
Therefore, for $\mbox{\it GIP\/}(x,y,z)$ as defined by Eqs.~(\ref{conds}) 
and (\ref{fcn}), the classical communication complexity is at most four 
and the quantum communication complexity is at most three.
Also, in \cite{CB}, it is shown that, in a slightly different communication 
model, the classical communication complexity of $\mbox{\it GIP\/}(x,y,z)$ 
is at least three.
A classical lower bound of four in our current communication model can be 
obtained by slightly modifying the proof in \cite{CB}.

\section{A two-party probabilistic scenario}

Consider the following probabilistic two-party communication complexity 
scenario.
Alice and Bob receive $x$ and $y$ respectively, 
where $x,y \in \{0,1\}^2$.
The common goal is to compute the value of the function 
\begin{equation}
g(x,y) = x_1 \xor y_1 \xor (x_0 \wedge y_0), 
\label{andxor}
\end{equation}
with as high probability as possible.
An execution is considered successful if and only if the value determined 
by Alice and the value determined by Bob are {\em both\/} correct.

In the following two subsections, we show that, with a prior quantum 
entanglement and two bits of communication, the probability of 
success can be at least $\cos^2({\pi \over 8}) = 0.853...$, whereas, 
with a shared random string instead of quantum entanglement, and two 
bits of communication, the probability of success cannot exceed $0.75$.
Thus, without prior entanglement, to achieve a success probability of 
at least $\cos^2({\pi \over 8})$, {\em three} bits of communication 
are necessary.

\subsection{With quantum entanglement}

We now show that if Alice and Bob initially share a 
certain entanglement of two qubits then there is a two-bit 
protocol in which both parties output the correct value of 
$g(x,y)$ with probability $\cos^2({\pi \over 8}) = 0.853...$.
The entanglement is a so-called Einstein-Podolsky-Rosen (EPR) pair 
\begin{equation}
\ket{\Q_A \Q_B} = 
{\textstyle{ 1 \over \sqrt 2}} (\ket{00} - \ket{11}).
\label{epr}
\end{equation} 

The idea is based on applying the result of Clauser, Horne, Shimony, 
and Holt \cite{CHSH} to enable Alice and Bob to obtain bits $a$ and $b$ 
such that 
\begin{equation}
\Pr[a \xor b = x_0 \wedge y_0] = \cos^2(\textstyle{\pi \over 8}).
\label{chsh}
\end{equation}
This is achieved by the following procedures:
\begin{tabbing}
\vspace*{8mm} Pr\=oce\=dure  for Alice: \hspace{5cm} Pr\=oce\=dure for Bob:\\
\> {\bf if} $x_0 = 0$ {\bf then} \> 
\> {\bf if} $y_0 = 0$ {\bf then} \\ 
\> \> {\bf apply} $R(-{\pi \over 16})$ {to} $\Q_A$ 
\> \> {\bf apply} $R(-{\pi \over 16})$ {to} $\Q_B$ \\
\> {\bf else} \> 
\> {\bf else} \\
\> \> {\bf apply} $R({3 \pi \over 16})$ {to} $\Q_A$ 
\> \> {\bf apply} $R({3 \pi \over 16})$ {to} $\Q_B$ \\
\> {\bf measure} $\Q_A$ {\bf yielding bit $a$} \> 
\> {\bf measure} $\Q_B$ {\bf yielding bit $b$} \\
\end{tabbing}

\ni In the above, $R(\theta)$ is the rotation by angle $\theta$, which is 
represented in the standard basis as 
\begin{equation}
R(\theta) = 
\pmatrix{\cos \theta & - \sin \theta \cr \sin \theta & \ \ \cos \theta},
\end{equation}
and the measurements are performed in the standard basis.

The fact that the above protocol satisfies Eq.~(\ref{chsh}) follows 
from the fact that if $R(\theta_1) \ox R(\theta_2)$ is applied to 
state (\ref{epr}) then the resulting state is 
\begin{equation}
\ket{\Q_A \Q_B} = 
{\textstyle{ 1 \over \sqrt 2}} 
\left(
\cos(\theta_1 + \theta_2) (\ket{00} - \ket{11}) + 
\sin(\theta_1 + \theta_2) (\ket{01} + \ket{10})
\right),
\end{equation}
which is straightforward to verify.

After this procedure, Alice sends $(a \xor x_1)$ to Bob, 
and Bob sends $(b \xor y_1)$ to Alice.
At this point, each party can determine the bit 
\begin{eqnarray}
(a \xor x_1) \xor (b \xor y_1) & = & x_1 \xor y_1 \xor (a \xor b),
\end{eqnarray}
which equals $x_1 \xor y_1 \xor (x_0 \wedge y_0)$ with probability 
$\cos^2({\pi \over 8})$, as required.

\subsection{With shared classical random bits but no quantum entanglement}

We now show that if Alice and Bob initially share classical random 
bits but no quantum entanglement then there is no two-bit protocol in 
which both parties output the correct value of $g(x,y)$ with probability 
greater than $3 \over 4$.
By Theorem 3.20 of \cite{KN}, it is sufficient to prove the lower bound on 
the error probability for all {\em deterministic} protocols with respect to 
{\em random inputs} from $\01^2 \x \01^2$ (which we can take to be uniformly 
distributed).
As noted in Section \ref{section:lowerbound}, we can represent any 
$2$-bit protocol as a binary tree of depth $2$, with non-leaf nodes 
labelled A(lice) and B(ob).

Assume, without loss of generality, that the root of the protocol-tree 
is labelled A.
The first bit that Alice sends is some function 
$\phi : \01^2 \rightarrow \01$ of her input data $x$ alone.
The function $\phi$ partitions $\01^2$ into two classes 
$S_0 = \phi^{-1}(0)$ and $S_1 = \phi^{-1}(1)$.
Let the first child and second child of the root correspond to the 
paths traversed when the first bit sent (by Alice) indicates that 
$x \in S_0$ and $x \in S_1$, respectively.
We must consider all partitions $S_0$ and $S_1$ in combination with all 
cases where the two children of the root are BB, AB, or AA (the case 
BA can be omitted by symmetry).\ee

\begin{table}
\begin{center}
\begin{tabular}{c|cccc|c}
\multicolumn{6}{c}
{\hspace*{2.5mm} 00\hspace*{1mm} 01\hspace*{1mm} 10\hspace*{1mm} 11} \\
\cline{2-5}
00 &  0 &  0 &  1 &  1 & \\
01 &  0 &  1 &  1 &  0 & \\
10 &  1 &  1 &  0 &  0 & \\
11 &  1 &  0 &  0 &  1 & \\
\cline{2-5}
\end{tabular}
\caption{\protect \small The values of $g(x,y)$.
The columns are indexed by $x$ and the rows are indexed by $y$.}
\end{center}
\end{table}

\loud{Lemma 3:}{\sl 
If the child corresponding to $S_i$ is labelled\/} B {\sl then, conditioned 
on $x \in S_i$, the probability that Bob correctly determines $g(x,y)$ 
is at most: $1$, if $|S_i| = 1$; $3 \over 4$, if $|S_i| = 2$; 
and $2 \over 3$, if $|S_i| = 3$.}\ee

\loud{Proof:}
The case where $|S_i| = 1$ is trivial.

For the case where $|S_i| = 2$, first consider the subcase where 
$S_i = \{00,01\}$.
Under the condition $x \in S_i$, $(x,y)$ is a position in one of the 
first two columns of the table, and Alice's bit to Bob indicates this 
to him.
{}From Bob's perspective, if $y = 00$ then $g(x,y) = 0$, so Bob can 
determine the correct answer.
Similarly, if $y = 10$ then $g(x,y) = 1$, so Bob can determine the 
correct answer.
However, if $y = 01$ then, since the first two columns of the table 
differ in this row, whatever function of Alice's message and $y$ 
Bob computes, the probability that it will match $g(x,y)$ is at most $\half$.
Similarly, if $y = 01$ then Bob computes the correct answer with 
probability at most $\half$.
Since these four values of $y$ are equiprobable, the probability that 
Bob correctly computes $g(x,y)$ conditioned on $x \in S_i$ is at most 
${1 \over 4} \cdot {1} + {1 \over 4} \cdot {1} + 
{1 \over 4} \cdot {1 \over 2} + {1 \over 4} \cdot {1 \over 2} = {3 \over 4}$.
The other five subcases in which $|S_i| = 2$ are handled similarly.

For the case where $|S_i| = 3$, first consider the subcase where 
$S_i = \{00,01,10\}$.
Under the condition $x \in S_i$, $(x,y)$ is a position in one of the 
first three columns of the table, and Alice's bit to Bob indicates this 
to him.
By looking at these three columns of the table, we observe that, 
from Bob's perspective, whatever the value of $y$, the probability 
of Bob determining $g(x,y)$ is at most $2 \over 3$.
The other two subcases in which $|S_i| = 3$ are handled similarly.
$\Box$\ee

Now, by Lemma 3, if the two children of the root are BB then the 
probability that Bob correctly determines $g(x,y)$ is at most: 
${1 \over 4} \cdot 1 + {3 \over 4} \cdot {2 \over 3} 
= {3 \over 4}$, 
if $|S_0| \neq |S_1|$; and 
${1 \over 2} \cdot {3 \over 4} + {1 \over 2} \cdot {3 \over 4} 
= {3 \over 4}$, 
if $|S_0| = |S_1|$.

Next, we show that, for protocol-trees in which the two children of 
the root are {\em not\/} BB, the correctness probability is actually 
less than $3 \over 4$.\ee

\loud{Lemma 4:}{\sl 
If the child corresponding to $S_i$ is labelled\/} A {\sl then, conditioned 
on $x \in S_i$, the probability that Alice correctly determines $g(x,y)$ 
is at most $1 \over 2$.}\ee

\loud{Proof:}
If the condition $x \in S_i$ occurs then Alice receives no information 
from Bob.
Therefore, from Alice's perspective, the value of $g(x,y)$ is either 
$y_1$, $y_1 \xor y_0$, $1 \xor y_1$, or $1 \xor y_1 \xor y_0$ (corresponding 
to the cases $x = 00$, $01$, $10$, and $11$ respectively).
The result now follows from the fact that, from Alice's perspective, 
$y$ is uniformly distributed over $\01^2$.
$\Box$\ee

By Lemma 4, it follows that, if the two children of the root are AA then 
the probability that Bob correctly determines $g(x,y)$ is at most 
${1 \over 2}$.
The remaining case is where the two children of the root are AB.
By applying Lemma 4 for the first child and Lemma 3 for the second child, 
the probability that both Alice and Bob correctly determine $g(x,y)$ is at 
most:
\begin{itemize}
\item
${1 \over 4} \cdot {1 \over 2} + {3 \over 4} \cdot {2 \over 3} 
= {5 \over 8}$,  
\ if $|S_0| = 1$ and $|S_1| = 3$ 
\item
${1 \over 2} \cdot {1 \over 2} + {1 \over 2} \cdot {3 \over 4} 
= {5 \over 8}$,  
\ if $|S_0| = 2$ and $|S_1| = 2$ 
\item
${3 \over 4} \cdot {1 \over 2} + {1 \over 4} \cdot {1} 
= {5 \over 8}$,  
\ if $|S_0| = 3$ and $|S_1| = 1$.
\end{itemize}

This completes the proof that no two-bit protocol is correct with 
probability more than $3 \over 4$.
There is a straightforward errorless three-bit protocol.

\section{Acknowledgments}

We would like to thank Charles Bennett, Lance Fortnow, Richard Jozsa, 
and Lev Vaidman for helpful discussions.

\end{document}